\documentclass[twocolumn]{aastex631}

\usepackage{ amssymb }
\usepackage{wrapfig}

\def\iso#1#2{\mbox{${}^{#2}{\rm #1}$}}
\def\fe6#1{\iso{Fe}{6#1}}
\def\mn5#1{\iso{Mn}{5#1}}
\def\al2#1{\iso{Al}{2#1}}
\def\pu24#1{\iso{Pu}{24#1}}

\def\pfrac#1#2{\left( \frac{#1}{#2} \right)}

\begin{document}

\title{Nearby Supernova and Cloud Crossing Effects on the Orbits of Small Bodies in the Solar System}

\author{Leeanne Smith}
\affiliation{Department of Atmospheric Sciences, University of Illinois, Urbana, IL 61801, USA}

\author[0000-0001-5071-0412]{Jesse A. Miller}
\affiliation{Department of Astronomy, Boston University, Boston, MA 02215, USA}
\affiliation{Department of Astronomy, University of Illinois, Urbana, IL 61801, USA}

\author[0000-0002-4188-7141]{Brian D. Fields}
\affiliation{Department of Astronomy, University of Illinois, Urbana, IL 61801, USA}

\begin{abstract}
Supernova blasts envelop many surrounding stellar systems, transferring kinetic energy to small bodies in the systems. Geologic evidence from \fe60 points to recent nearby supernova activity within the past several Myr. Here, we model the transfer of energy and resulting orbital changes from these supernova blasts to the Oort Cloud, the Kuiper belt, and Saturn's Phoebe ring. For the Oort Cloud, an impulse approximation shows that a 50 pc supernova can eject approximately half of all objects less than 1 cm while altering the trajectories of larger ones, depending on their orbital parameters. 
For stars closest to supernovae, objects up to $\sim100$ m can be ejected.
Turning to the explored solar system, we find
that supernovae closer than 50 pc may affect Saturn's Phoebe ring and can sweep away Kuiper belt dust. It is also possible that the passage of the solar system through a dense interstellar cloud could have a similar effect; a numerical trajectory simulation shows that the location of the dust grains and the direction of the wind (from a supernova or interstellar cloud) has a significant impact on whether or not the grains will become unbound from their orbit in the Kuiper belt. Overall, nearby supernovae sweep micron-sized dust from the solar system, though whether 
the grains are ultimately
cast towards the Sun or altogether ejected depends on various factors. Evidence of supernova-modified dust grain trajectories may be observed by {\em New Horizons}, though further modeling efforts are required.

\end{abstract}


\section{Introduction} 
\label{sec:intro}

Measurements of live \fe60 ($t_{1/2} = 2.6$ Myr) in geological deposits show that Earth was in the vicinity of supernovae (SNe) $\sim 3$ Myr and 
$\sim$ 7 Myr ago \citep{knie_indication_1999, knie_60fe_2004, fitoussi_search_2008, ludwig_time-resolved_2016, wallner_recent_2016, wallner_60fe_2021}. Freshly synthesized \fe60 from the explosion was incorporated into dust grains, which propagated through interstellar space and the solar system, eventually landing on Earth \citep{athanassiadou_penetration_2011, fry_magnetic_2020}.

While many studies have investigated the effects of supernovae on the Earth \citep[e.g.,][]{ellis_geological_1996, gehrels_ozone_2003, thomas_terrestrial_2016}, there have been far fewer investigations into effects throughout the solar system. \citet{stern_influence_1988} and \citet{stern_ism-induced_1990} calculated how SNe could flash heat the surface of Oort Cloud bodies and how dynamical drag could erode their surfaces. Armed with the evidence of recent supernova activity, we are motivated to consider a wider variety of ways SNe may have affected our solar system. In this work, we consider the possibility that nearby SNe may have dynamically altered the orbits of small solar system bodies.

The Oort Cloud, being distant from the Sun and therefore only tenuously gravitationally bound \citep{oort_structure_1950}, keenly experiences 
extrasolar
influences. These influences include the Galactic tidal field \citep{heisler_influence_1986} and nearby passing stars \citep{rickman_injection_2008}. In addition, the gravitational perturbation of passing giant interstellar clouds has been investigated \citep{jakubik_dynamics_2008, hut_have_1985}, though not the direct gasdynamical impulse of these clouds.

Recently, \citet{opher_possible_2024} proposed that the solar system travelled through a dense cold cloud called the Local Lynx of Cold Cloud (LxCC) near the same time as the initial onset of the \fe60 pulse 3 Myr ago. The dense cloud ($n_{\rm H} \sim 3000$ cm$^{-3}$) would have compressed the heliosphere to a mere 0.22 au, stripping the entire solar system of its protective solar wind.

Both the direct SN blast wave and cold cloud scenarios call for a significant shift in ram pressure felt throughout the solar system. While a SN blast does not change the density of the local interstellar medium (ISM) greatly (at most a factor of 4 due to a strong shock), the speed increases by a factor of $\sim$10, depending on the distance to the SN. In contrast, the cold cloud enhances ISM density by 4-5 orders of magnitude with little change in velocity.

The SN blast wave and cold clouds scenarios also differ in the regions they influence. The cold cloud coats the entire solar system, even Mercury's orbit \citep{opher_possible_2024}. The effects of the SN blast, on the other hand, are highly dependent on the distance to the SN. Estimates for the distance to the 3 Myr-old SN vary depending on many factors such as the total \fe60 abundance generated by the SN, amount of \fe60 incorporated into dust grains, and \fe60 uptake into geological samples \citep{ertel_supernova_2023}. Based on \fe60 measurements, the SN distance was approximately $20-140$ pc from Earth \citep{ertel_distances_2023, fry_astrophysical_2015-1}. \citet{miller_heliospheric_2022} performed simulations of how the heliosphere responds to SNe at different distances, finding that a SN at a moderate distance of 50 pc compresses the heliosphere to 10 au, the orbit of Saturn. Therefore, the effects of a SN blast wave are not felt as widely throughout the solar system as those from a cold cloud. 
Observable traces of
the SN blast within the solar system could potentially be used as a way to indicate SN distance, or even distinguish which of these two cases occurred.

To this end, we investigate the effects of how a large ram pressure from the SN blast or cold cloud could influence solar system orbits. This examination is done in three cases: section \ref{sec:Oort} describes an impulse approximation for Oort Cloud bodies, and section \ref{sec:Saturn} considers the effects on Saturn's rings and Kuiper belt dust. Finally, discussion and conclusions are given in section \ref{sec:discuss}.

\section{Oort Cloud}
\label{sec:Oort}

\subsection{Impulse Approximation}
\label{subsec:Oort_approx}

Due to the significant difference between Oort Cloud orbital time (1 Myr at $10^4$ au) and duration of the SN blast in the Sedov solution ($\sim$0.05 Myr), the SN blast wave imparts an impulsive force onto the Oort Cloud. Rather than assuming a time-dependent force, it is valid to simply assume that the blast wave imparts an instantaneous velocity boost to the comets. Assuming the comet absorbs all incident kinetic energy from the blast, the comet gains a kick velocity of
\begin{equation}
\label{eqn:vkick}
    v_{\rm kick} = \sqrt{ { \frac{3}{8 \pi}} \frac{E_{SN}} {\rho_c r_{c} D^2}} ,
\end{equation}
where $E_{SN}$ is the energy of the supernova (assumed to be $10^{51}$ erg), $r_c$ and $\rho_c$ are the radius and density of the comet, and $D$ is the distance from the supernova.

This is in no way intended to be an exact description of the modern distribution of Oort Cloud comets, as such information is outside the scope of this paper.
Rather, our purpose is to describe what would happen to such comets when they are hit by a SN blast.

These simulations consist of $10^6$ Oort Cloud objects with radii ranging from $10\ \mu \textrm{m -- } 100$ m and densities ranging from $0.5 \textrm{--} 1.5 \textrm{ g cm}^{-3}$. The orbital parameters, as well as size and density, are all chosen independently. This wide radius range allows us to capture the dynamics of all objects from dust grains to modest asteroids. Consistent with observed comets, they are given highly eccentric orbits ($0.9 < e < 1$) with semi-major axes within $10,000 < a < 20,000$ au \citep{dones_origin_2015}. Each comet has randomly generated Keplerian coordinates, including the angle of inclination $i$, argument of periapsis $\omega$, right ascension of the ascending node $\Omega$, and mean anomaly $\nu$ within ranges of $[0, \pi)$, $[0, 2 \pi)$, $[0, 2 \pi)$, and $[0, 2 \pi)$ respectively.

Each comet was converted to Cartesian coordinates, given an additional $v_{kick}$ in the $v_x$ direction, and converted back to Keplerian coordinates to analyze orbital properties.
Note that the direction of the kick itself is unimportant, as the spherical nature of the Oort Cloud ensures a symmetric geometry, so results will be independent of direction.

Due to the uncertainty in the distance to the SN, we employ a set of calculations for four SNe at distances of 10, 20, 50, and 100 pc.

\subsection{Results}
\label{subsec:Oort_results}

\begin{figure}[!htb]
    \centering
    \includegraphics[width=0.49\textwidth]{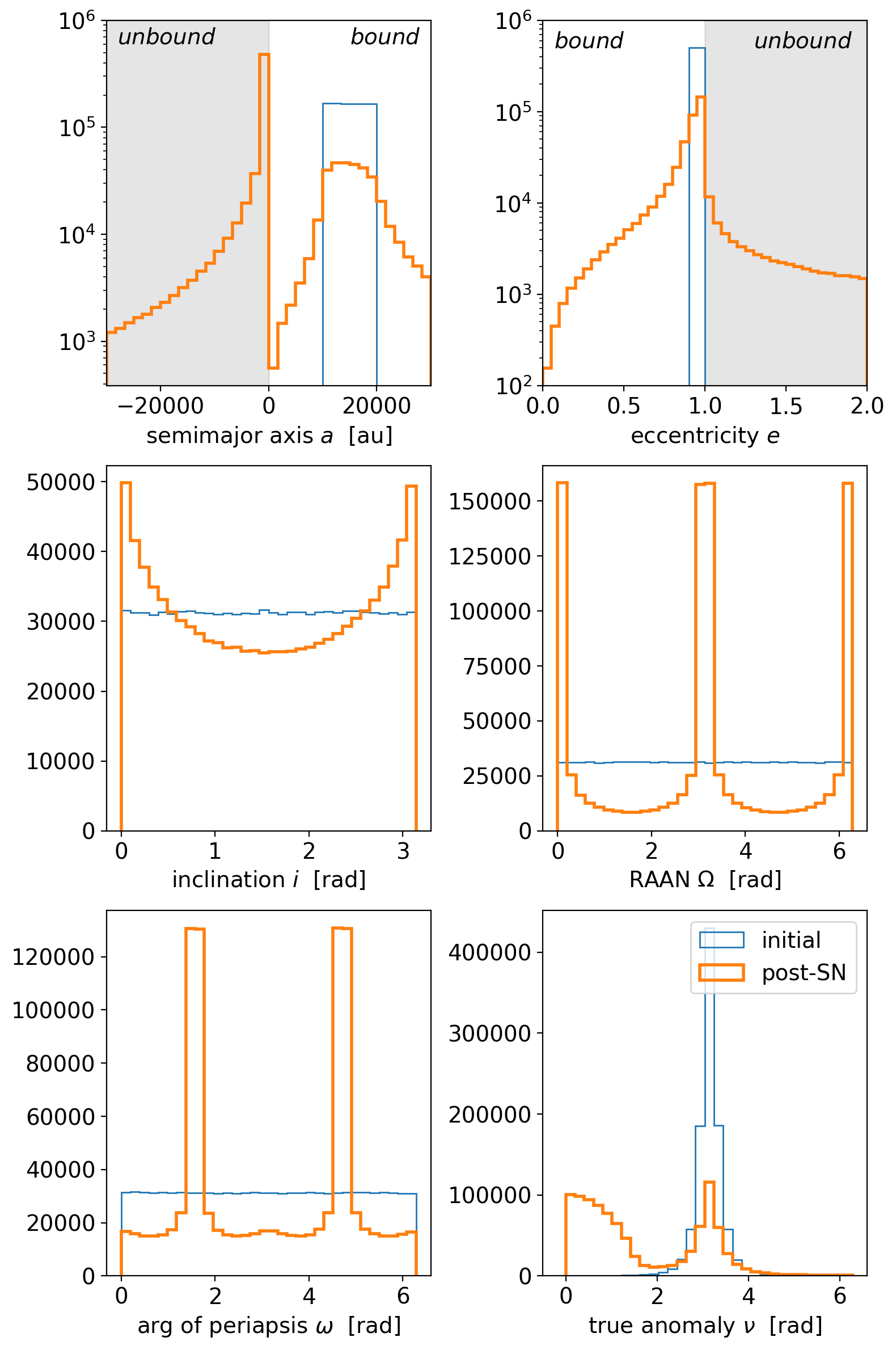}
    \caption{Histograms of orbital parameters of Oort Cloud comets in their initial distribution (blue) and after being hit by a remarkably close supernova 10 pc away (orange). The parameters shown are (left to right, top to bottom) semi-major axis, eccentricity, angle of inclination, right ascension of the ascending node, argument of periapsis, and the true anomaly.}
    \label{fig:OCorbits}
\end{figure}

The initial distribution and results of applying the kick velocity of a SN 10 pc distant are shown in Fig.~\ref{fig:OCorbits}.
For the semi-major axis, there are significantly fewer objects within the initial semi-major axis range because the rest were either pushed to a different orbit or completely ejected (which is treated as a negative semi-major axis). As will be shown later, the amount the semi-major axis changes is mostly determined by the body's mass and only weakly dependent on the orbital parameters.
The eccentricity reveals that while many bodies became more circularized, many others were unbound with $e \gg 1$ and $a < 0$.

The three angular coordinates ($i$, $\Omega$, and $\omega$) have more interesting features. In these plots, unlike those of semi-major axis and eccentricity, all orbits are represented. The inclination shows a slight preference to move towards $i=0$ and $\pi$. There is a strong preference to shift the right ascension towards $\Omega = 0$, $\pi$, and $2\pi$ and the argument of periapsis towards $\omega = \pi/2$ and $3\pi/2$. The large spikes are the result of becoming unbound: unbound bodies are the closest they get to the reference plane that defines $i=0$, and so the point at which they cross this plane (the right ascension) is near their current position. The argument of periapsis is strongly peaked because the unbound bodies are very near their periapsis and travelling in a near-straight line, which corresponds to being at the stated values of $\omega$. 
Lastly, the true anomaly is clustered towards 0 for similar reasons: unbound particles are very near their periapsis, which is defined as the point at which $\nu=0$.

\begin{figure}
\centering
    \includegraphics[width=0.48\textwidth]{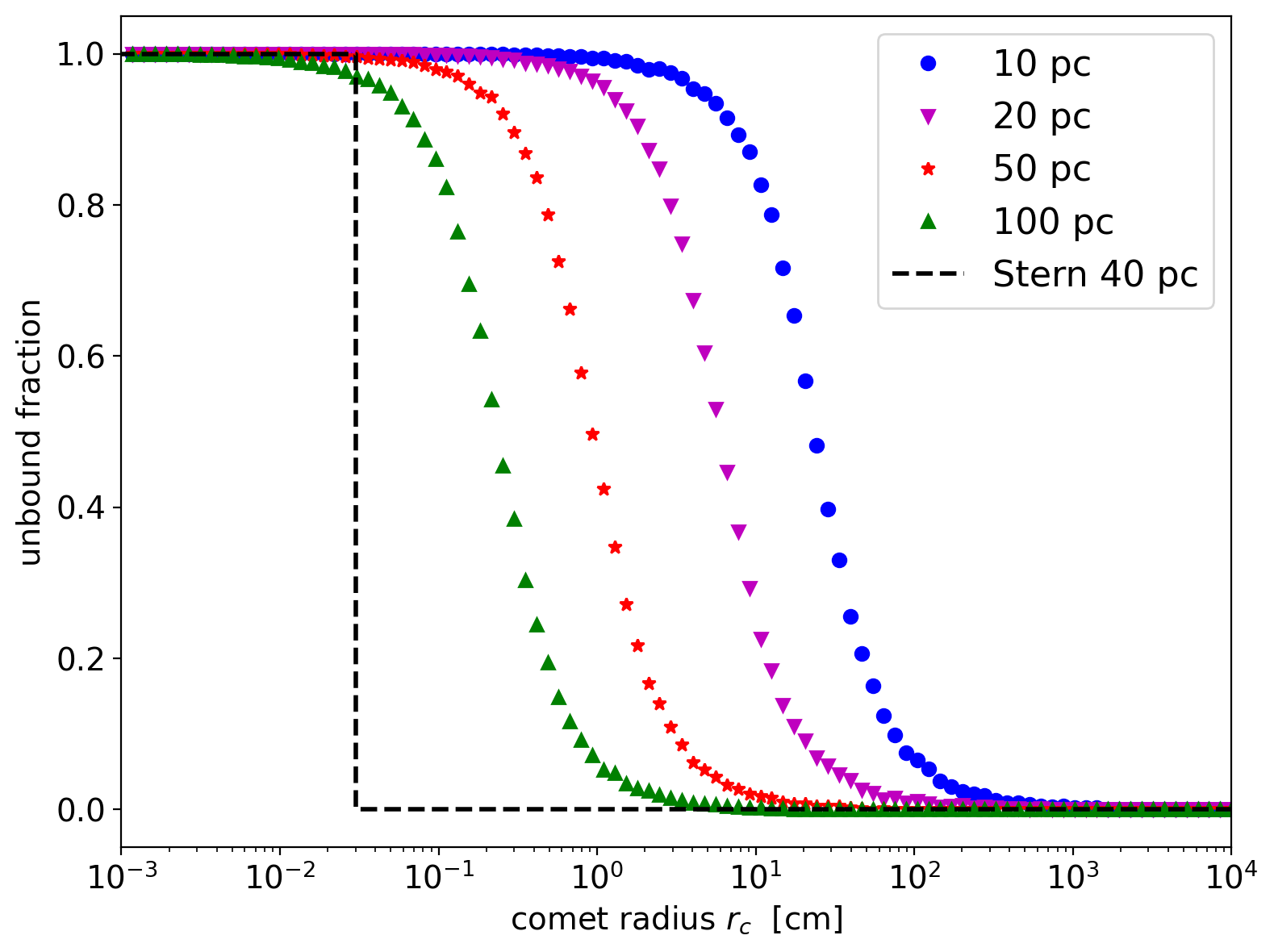}
    \caption{Fraction of comets of a given size that become unbound from a SN blast, evaluated for SN distances of 10, 20, 50, and 100 pc from the solar system. The dashed line comes from a simplified calculation by \citet{stern_ism-induced_1990} for a SN 40 pc away for particles orbiting 20,000 au from the Sun.}
    \label{fig:OCunbound}
\end{figure}

The simulated objects were sorted and grouped into logarithmically-spaced radius bins. For each bin, the percentage of orbits that became unbound was calculated. The result is shown in Figure \ref{fig:OCunbound}. As expected, the unbound fraction shows a strong dependence on size, with the lighter particles being completely ejected while larger ones barely move. This is also highly dependent on SN distance. For SNe at 10, 20, 50, and 100 pc away, we see steep drop offs in the number of unbound objects with radii around 20, 5, 0.8, and 0.2 cm respectively.

We also compare our results to those of \citet{stern_ism-induced_1990}, who performs a similar calculation, albeit without the complete description of 3D orbital mechanics utilized here. That study finds that all particles at $a > 10^4$ au with $r_c <$ 0.3 mm are ejected by a SN at 40 pc. We find a fairly good agreement for the particles that become completely unbound. However, there are a large number of particles that can become unbound due to their orbital parameters that \citet{stern_ism-induced_1990} does not account for.

We see that at distances appropriate for the recent supernovae ($\sim 50-100$ pc), Oort Cloud particles of $\sim 1 \ \rm cm$ will be strongly affected, with some ejected to interstellar space.  Other such particles will remain bound, in orbits coming much closer to the Sun; we have calculated the perihelion distribution of the perturbed particles and find that only a tiny fraction of the particles reach within 100 au.  For supernovae close enough to be dangerous (10-20 pc), objects as large as $\sim 1\ \rm m$ will suffer these effects.

\section{Saturn's Phoebe Ring and Kuiper Belt Dust}
\label{sec:Saturn}

Unlike in the Oort Cloud, where the $\sim 0.05$ Myr passage of a SN blast could be modelled as an impulse approximation, bodies that make up planetary rings and the Kuiper belt orbit much more rapidly. The total force imparted by the blast spans many orbits rather than a single instant. To investigate these orbital changes, a more complete model is required. Therefore, we make estimates using a secular approximation, followed by a numerical implementation.

Two cases stand out as particularly enlightening: Saturn's rings and Kuiper belt dust. Saturn's A and B rings have a short dynamical timescale and high particle densities, making collisions frequent. The collision timescale is much shorter than the timescale for the wind perturbations to become effective, so that these rings will be unaffected by the SN.\footnote{We thank the anonymous referee for pointing this out.} However, Saturn's largest ring, the Phoebe ring, has a much lower number density, and grain sizes are approximately 10 $\mu$m \citep{verbiscer_saturns_2009}, making the Phoebe ring a valid target for our calculation. Similarly, Kuiper belt dust represents tiny bodies that keenly experience an outside force, and their modified trajectories can be compared to spacecraft measurements today \citep{poppe_constraining_2019}.

\subsection{Secular Approximation: Formalism}

The trajectories of small grains in the solar system are influenced by a drag force,
\begin{equation}
    \label{eqn:F_drag}
    F_{\rm drag}(t) = C_{\rm d}\ \rho_{\rm w} \ \pi r_{\rm gr}^2 \ |\vec{v}_{\rm rel}|^2 \ \hat{v}_{\rm rel}
\end{equation}
Here, $\vec{v_{rel}} = \vec{v}_{\rm w} - \vec{u}$ gives the difference between the wind velocity $\vec{v}_{\rm w}$
and the particle velocity $\vec{u}$; this relative velocity has magnitude $|\vec{v}_{\rm rel}|$ 
and direction $\hat{v}_{\rm rel}$.
We take the drag coefficient $C_{\rm d} = 1$.
Finally, $\rho_{\rm w}$ is the wind density and $r_{\rm gr}$ is the grain radius.

This leads to a slow (secular) change of the orbital parameters for each particle.  \citet{pastor_orbital_2011} derived the expression governing these changes, time-averaged and to leading order in the perturbations. We focus on the case when $v_{\rm w}/u \gg 1$, which is valid for the high velocity of a SN blast. The semi-major axis evolves as
\begin{equation}
\label{eq:dadt}
    \frac{da}{dt}
    = - 2\ C_{\rm d}\ n_{\rm w} \frac{m_{\rm w}}{m_{\rm gr}} \pi r_{\rm gr}^2 v_{\rm w} \ a \ f
\end{equation}
where we have separated the wind density into $\rho_{\rm w} = m_{\rm w} n_{\rm w}$, and $m_{\rm gr}$ is the grain mass.
$f$ is a dimensionless function that depends on the eccentricity and angular orbital parameters, and the orientation of the wind relative to the orbital plane; in our estimate we will take it to be $f \sim 1$.
We see from eq.~(\ref{eq:dadt}) that the characteristic timescale for 
the change in the semi-major axis is
\begin{equation}
    \Gamma_a = \frac{|da/dt|}{a}  \ \sim \ 2\ C_{\rm d}\ n_{\rm w} \frac{m_{\rm w}}{m_{\rm gr}} \pi r_{\rm gr}^2 \frac{F_{\rm drag}}{m_{\rm gr} v_{\rm w}} 
\end{equation}
which we see is just twice the ratio of the drag force 
on a grain divided by the grain momentum $m_{\rm gr} v_{\rm w}$ as seen by the wind.
This gives a characteristic decay timescale
\begin{eqnarray}
    \label{eqn:tau_a}
    \tau_a & = & \frac{1}{\Gamma_a}  \\ 
    & = & 5 \ {\rm Myr} \pfrac{500\ {\rm km/s}}{v_{\rm w}} \pfrac{10^{-2}\ {\rm cm}^{-3}}{n_{\rm w}}
    \pfrac{1 \ \mu{\rm m}}{r_{\rm gr}}^2
\end{eqnarray}
for a grain of density $\rho_{\rm gr}=2 \ \rm g/cm^3$,
which is much longer than the $\sim 0.05 \ \rm Myr$ timescale 
a SN blast will extend to the location of Saturn or the Kuiper belt \citep{miller_heliospheric_2022}.

For the grain orbital eccentricity, \citet{pastor_orbital_2011} find the time-averaged change to be
\begin{eqnarray}
    \frac{de}{dt} & = & -C_{\rm d} n_{\rm w} \frac{m_{\rm w}}{m_{\rm gr}} v_{\rm w} \left( g_1 + \frac{v_{\rm w}}{v_c} g_2 \right) \\
    & = & -\frac{1}{2} \Gamma_a  \left( g_1 + \frac{v_{\rm w}}{v_c} g_2 \right)
\end{eqnarray}
where the dimensionless factors $g_1$ and $g_2$ are functions of the eccentricity, angular orbital parameters, and wind orientation.  We see that there are two terms,
the first of which has the same order of magnitude as $\Gamma_a$, and thus will change over the same timescale.
The second term differs by the ratio of the wind speed to the circular orbit speed $v_c = \sqrt{GM/a}$.
We see that the eccentricity decay rate is of order
\begin{equation}
\Gamma_e \sim \frac{1}{2} \frac{v_{\rm w}}{v_c} \Gamma_a 
\end{equation}
and thus the timescale for eccentricity change due to this term is
\begin{equation}
    \label{eqn:tau_e}
    \tau_e = \frac{1}{|de/dt|} \sim 2 \frac{v_c}{v_{\rm w}} \tau_a
\end{equation}
and finally,
\begin{eqnarray}
    \label{eqn:tau_ea_ratio}
    \frac{\tau_e}{\tau_a} & = & 2 \frac{v_c}{v_{\rm w}}  \\
    & \sim & 0.02 \ \pfrac{500\ {\rm km/s}}{v_{\rm w}} \pfrac{M}{M_\odot}^{1/2} \pfrac{40\ {\rm au}}{a}^{1/2}
\end{eqnarray}
and we see that for SN conditions, the eccentricity will change more rapidly than the semi-major axis (eq.~\ref{eqn:tau_a}). 
For the semi-major axis in eq. \ref{eqn:tau_a}, this amounts to $\tau_e \sim 0.1$ Myr.

The angle of inclination $i$ has a similar form as the eccentricity and so may be treated in a similar manner. This rate of change and corresponding timescale, then, are
\begin{eqnarray}
    \frac{di}{dt} & = & - \frac{1}{4} \Gamma_a \left( h_1 + \frac{v_{\rm w}}{v_c} h_2 \right) \\
    \label{eqn:tau_i}
    \tau_i & \sim & 4 \frac{v_c}{v_{\rm w}} \tau_a
\end{eqnarray}
where $h_1$ and $h_2$ are defined in the same manner as $g_1$ and $g_2$. The inclination timescale is merely a factor of 2 larger than the eccentricity timescale.

\subsection{Secular Approximation: Results}

Equations (\ref{eqn:tau_a}), (\ref{eqn:tau_e}), and (\ref{eqn:tau_i}) can be used to derive the timescales on which orbital changes to Saturn's Phoebe ring and Kuiper belt dust take place. Given that these equations do not account for geometrical factors, these timescales should not be taken too precisely. These timescales are shown in Table \ref{tab:runs}. Note that they must be compared to the duration of the wind rather than the time since the event.
For Saturn's Phoebe ring, we assume a 10 $\mu$m ring particle with a density of 1.6 g/cm$^3$ and semi-major axis 150 $R_{\rm Saturn}$ \citep{verbiscer_saturns_2009}, whereas a Kuiper belt dust grain has a size of 1 $\mu$m and a density of 2 g/cm$^3$ at 40 au. The SN density and velocity are calculated with a Sedov-Taylor profile in an ambient medium of 0.005 cm$^{-3}$, taken from \citet{miller_heliospheric_2022} and showing the onset of the blast. While the distance to the SN 3 Myr ago is uncertain, it is likely in the range of 20-140 pc based on \fe60 measurements \citep{ertel_distances_2023,fry_astrophysical_2015-1}. We include a 20 pc case here, representing an extreme event.
In addition, since the incoming wind is completely governed by density and velocity, we include the scenario of passing through the LxCC, as proposed by \citet{opher_possible_2024}.

\begin{table*}[]
    \label{tab:runs}
    \caption{Parameters for Ring Simulations Timescales for Orbit Perturbations}
    \begin{tabular}{c|ccc|ccc}
        \hline \hline
       Case & Wind density & Wind velocity & System & $a$ Timescale & $e$ Timescale & $i$ Timescale \\
       & $n \ [\rm atoms/cm^3]$ & [km/s] & & $\tau_a \ [\rm yr]$ & $\tau_e \ [\rm yr]$ & $\tau_i \ [\rm yr]$ \\
       \hline
        SN, 100 pc & 0.02 & 273 & Phoebe ring & 2.8e7 & 4.3e5 & 8.7e5 \\
        SN, 100 pc & 0.02 & 273 & Kuiper Belt & 1.8e6 & 6.1e4 & 1.2e5 \\
        SN, 50 pc & 0.02 & 771 & Phoebe ring & 1.0e7 & 5.4e4 & 1.1e5 \\
        SN, 50 pc & 0.02 & 771 & Kuiper Belt & 6.3e5 & 7.7e3 & 1.5e4 \\
        SN, 20 pc & 0.02 & 3048 & Phoebe ring & 2.5e6 & 3.5e3 & 7.0e3 \\
        SN, 20 pc & 0.02 & 3048 & Kuiper Belt & 1.6e5 & 490 & 980 \\
        LxCC & 3000 & 14.1 & Phoebe ring & 3.7e3 & 1.1e3 & 2.2e3 \\
        LxCC & 3000 & 14.1 & Kuiper Belt & 230 & 150 & 310 \\
        \hline \hline
    \end{tabular}
\end{table*}

We find that the timescale for a SN to alter the orbit of particles in Saturn's Phoebe ring is $10^4 - 10^5$ years for the eccentricity and inclination, though the semi-major axis takes longer. A supernova blast must be closer than $\sim50$ pc to compress the heliosphere enough to reach Saturn in the first place \citep{miller_heliospheric_2022, tamayo_radial_2016}. The exposure time would be in the range of 1-10 kyr, in surprising agreement with the eccentricity and inclination timescales calculated here. Therefore, we conclude that nearby supernovae may in fact affect the tilt and shape of the Phoebe ring. Passing through a dense interstellar cloud as in \citet{opher_possible_2024} requires an exposure time of about 1 kyr. Given the uncertainties in the size of the LxCC, it is unclear whether or not this timescale could be achieved.

The results are not drastically different for Kuiper belt dust.
But in this case, the SN influence occurs over 0.1-1 Myr, depending on the distance. When comparing with the exposure time from \citet[][their Fig.~11]{miller_heliospheric_2022}, we see that such timescales are typically around 50 kyr at a distance of 40 au, nearly independent of SN distance. A nearby SN may indeed alter the orbit of Kuiper belt dust grains. While the semi-major axis is not shifted greatly, the eccentricity and inclination have timescales two orders of magnitude less, and thus SNe will still modify their orbits in interesting ways. 
In their new orbits,
the grains would experience additional effects like the gravitational tug of other planets.

The passage through dense interstellar clouds would certainly have a large effect of Kuiper belt dust. The orbit decay timescale $\tau_a \sim 200$ yr for the LxCC is short, and so grains are significantly affected over even a single orbit. In the SN cases, $\tau_a$ was two orders of magnitude greater than $\tau_e$ and $\tau_i$. However, because the LxCC has such a lower velocity, eq.~(\ref{eqn:tau_ea_ratio}) implies that this difference in velocity is what causes the timescales to be similar here. Such an order of magnitude calculation suggests that this scenario deserves a closer calculation.

\subsection{Numerical Integration}

Given the surprisingly short timescales to modify the orbits of Kuiper belt dust, a numerical simulation is in order.
The behavior of the particles is influenced both by the drag force created by the blast wave (eq. \ref{eqn:F_drag}) and Newtonian gravity. A more complete model, such as those of e.g., \citet{slavin_trajectories_2012} or \citet{poppe_improved_2016}, would account for additional forces like magnetic fields or Poynting-Robertson drag. We leave such efforts for future work.
The equation of motion for the particle of velocity $\vec{u}$ is
\begin{equation}
    \label{eqn:force}
    m_\mathrm{gr} \frac{d\vec{u}}{dt} = -(1-\beta) GM_\odot m_\mathrm{gr} \frac{\vec{r}}{|\vec{r}|^3} + F_{\rm drag},
\end{equation}
which we integrate over several orbital periods. 
Here $\beta$ is the ratio of radiation pressure force on the grain to gravity;
$\beta < 1$ gives a net attractive force for grains that orbit the Sun in the absence of a wind.
In our numerical results we will take $\beta = 0$, but for nonzero values
the timescale would increase by a factor $(1-\beta)^{-1/2}$.

\begin{figure*}
    \centering
    \includegraphics[width=0.9\textwidth]{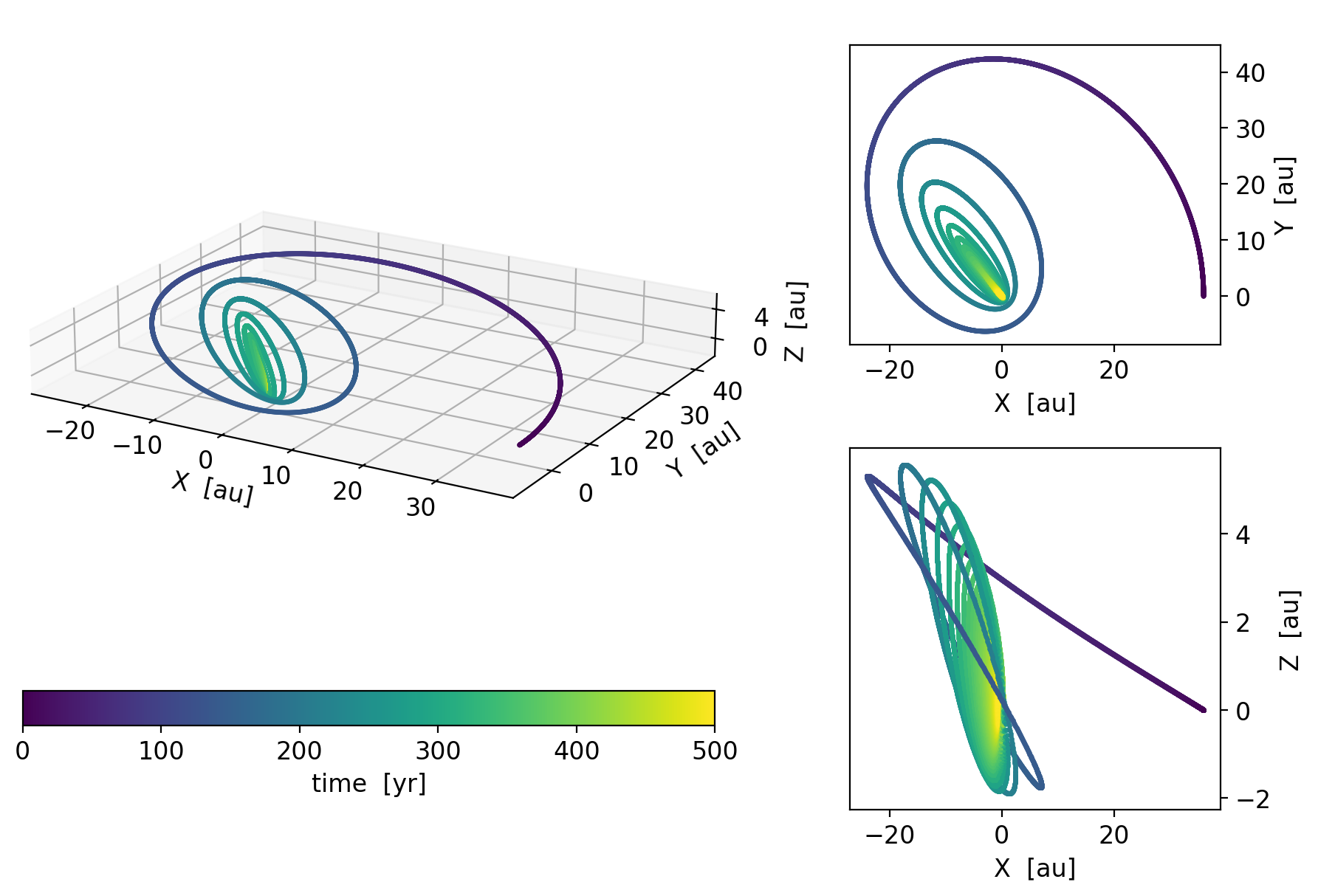}
    \caption{One possible trajectory of a 1 $\mu$m dust grain in the Kuiper belt subject to a constant wind from a dense interstellar cloud, corresponding to the ``LxCC'' case in Table \ref{tab:runs}. (a) shows the full 3D path, and (b) and (c) show projections in the $x$-$y$ and $x$-$z$ planes, respectively. Note that (c) is not an equal aspect ratio.}
    \label{fig:3D-orbit}
\end{figure*}

As an example, Fig.~\ref{fig:3D-orbit} shows the orbit of a 1 $\mu$m dust grain subject to the high density of the LxCC (the ``LxCC Kuiper Belt'' case in Table \ref{tab:runs}). Initially, the grain has a semi-major axis of 40 au and an eccentricity of 0.1, and the wind blows in the [$\hat{x}$, $\hat{y}$, $\hat{z}$] direction, which was chosen to give an oblique angle to the ecliptic and to give bound trajectories. Over a duration of 500 years, the orbit changes drastically. The most apparent change is the semi-major axis, which has diminished to less than 1 au by the end. Also noteworthy is the inclination, which has been altered so greatly the particle orbits nearly perpendicular to its initial orbit. Significant changes occur on the order of $\sim100$ years, in good agreement with the secular approximation. In this instance, the dust grain remained bound in the solar system; however, by merely changing the direction of the wind, it is readily possible to eject the grain. For example, if the wind comes from the ecliptic pole, the grain is unbound from the system.  For $\mu$m-sized dust, then, even the direction of the wind has a significant effect on the fate of Kuiper belt dust. But in all scenarios the grain no longer orbits in the Kuiper belt.

\begin{figure*}
    \centering
    \includegraphics[width=0.9\textwidth]{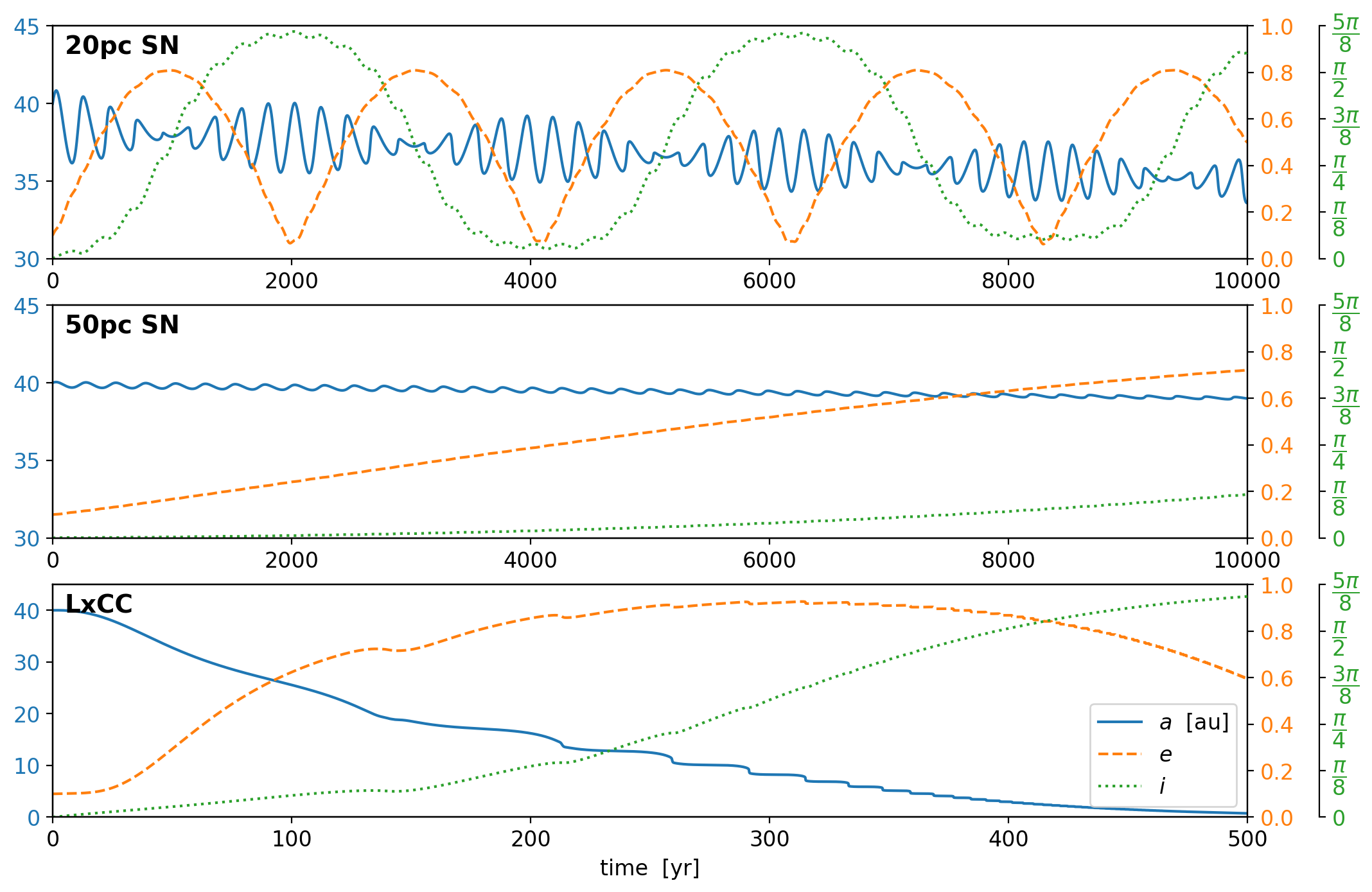}
    \caption{Integrated time-evolution of a 1 $\mu$m Kuiper belt dust grain subject to a wind from a 20 pc SN (top), a 50 pc SN (middle), and the LxCC (bottom). Plotted quantities are the semi-major axis (blue solid), eccentricity (orange dashed), and inclination (green dotted). Note that in the bottom panel, the semi-major axis and time are on different scales than the top and middle.}
    \label{fig:orbit_evolution}
\end{figure*}

This numerical exercise is repeated for two SN scenarios, a 20 pc and 50 pc case. They are given the same initial conditions as in the LxCC scenario, but are integrated over a longer time to be consistent with the duration of a SN passage \citep{miller_heliospheric_2022}. The evolution of the semi-major axes, eccentricities, and inclinations are shown in Fig.~\ref{fig:orbit_evolution}. As expected from the secular approximation, the eccentricity changes rapidly while the long-term evolution semi-major axis occurs over much longer timescales.

These numerical simulations, in addition to validating the timescale estimates of the secular approximation, show exactly how dust grain orbits react to external winds. Further study is needed to apply this process to the observed distribution of interplanetary dust, which will show how the entire dust distribution changes.

\section{Discussion \& Conclusions}
\label{sec:discuss}

We investigated how a nearby SN blast can dynamically alter the orbits of bodies in our solar system. In the first case, we employed an impulse approximation to model the effects on Oort Cloud bodies. By allowing for a wide range of radii, we found that a SN 50 pc distant will leave most $\gtrsim 10$ cm bodies unaltered in their orbit, but ejects all dust grains smaller than 1 mm from the solar system. From the geological \fe60-based rate of $\sim 2$ SNe per 10 Myr, these events would sweep clear the Oort Cloud of small debris many times over the age of the solar system. Shortly after the event, debris will be replaced as collisions occur and larger bodies erode. \citet{stern_influence_1988} has also shown that another effect of nearby SNe is to flash-heat icy bodies, which could potentially alter their surface properties.

In addition to our own solar system, a roughly spherical SN blast will also collide with all nearby stellar systems. With a similar stellar density to the solar neighborhood, the nearest of these stars will be $\sim$1 pc distant. If that star hosts an exo-Oort Cloud, then the supernova blast could eject objects of up to $\sim 100$ m into interstellar space, seeding the Milky Way with interstellar comets and asteroids. 
Thus, SN ejection represents a possible mechanism to create objects like 'Oumuamua \citep{meech_brief_2017}.

A secular approximation is applied to the orbits of Saturn's Phoebe ring and the Kuiper belt. The timescale for a SN to significantly alter Saturn's A and B rings is extremely large, but that of the Phoebe ring is surprisingly within reach for a SN within 50 pc. For a dense interstellar cloud, this timescale is on the order of $\sim$kyr.
If our solar system crossed the LxCC, it would have had a minor effect on the Phoebe ring. However, crossings of GMCs typically take 0.1-1 Myr \citep{talbot_encounters_1977}, and so would certainly devastate the Phoebe ring. While crossing GMCs with a density high enough to expose Earth statistically occur every $\sim$Gyr, the density required to expose Saturn (assuming ram pressure balance) is $(a_{\rm S} / a_\oplus)^2 = 90$ times less, and should occur much more frequently.

Based on the same secular approximation, the distribution of Kuiper belt dust exposed to a SN blast may be significantly altered. If the ejection fraction were roughly similar to that of the Oort Cloud, a supernova 50 pc distant (such as the one 3 Myr ago) would have ejected all particles less than $\sim$1 mm from the Kuiper belt. This result would indicate that the dust observed by \textit{New Horizons} \citep{poppe_constraining_2019} has been generated in the past 3 Myr. Indeed, such dust may not have reached an equilibrium state, as the Poynting-Robertson drag time is $\sim$5 Myr \citep{moro-martin_dynamical_2003, wyatt_poynting-robertson_1950}. In addition, the observed Kuiper belt dust production rate of $10^7$ g s$^{-1}$ \citep{poppe_improved_2016} and a total dust mass of $3.5 \times 10^{18}$ kg \citep{poppe_constraining_2019} lead to a dust ``replenishment'' timescale of 11 Myr. Both these Poynting-Robertson and replenishment timescales are longer than the age of the most recent SN blast, so it is possible that echoing effects still exist in dust grain trajectories today. Furthermore, recent reports from the \textit{New Horizons} spacecraft have shown an unexpected increase in dust flux \citep{doner_new_2024}. Whether this increase in dust flux is in some way related to recent interstellar factors is left for future study. These efforts may determine whether extant effects in Kuiper belt dust may bear the scars of ancient SNe.
 
\begin{acknowledgments}
We gladly acknowledge many enlightening discussions with Adrienne Ertel, Erica Albrigo, Merav Opher, and Alex Doner. We are particularly indebted to Scott Tremaine for discussions which inspired this work.  We also thank the Undergraduate Research Apprenticeship Program (URAP) at UIUC. JAM acknowledges support from the NASA/DRIVE program entitled ``Our Heliospheric Shield'', 80NSSC22M0164, \url{https://shielddrivecenter.com}.
The work of BDF  was supported by the NSF under grant number AST-210858.
\end{acknowledgments}

\vspace{5mm}
\software{astropy \citep{robitaille_astropy_2013, price-whelan_astropy_2018},
          numpy \citep{harris_array_2020},
          matplotlib \citep{hunter_matplotlib_2007},
          scipy \citep{virtanen_scipy_2020}
          }

\bibliography{SN_SolarSystem}{}
\bibliographystyle{aasjournal}

\end{document}